\documentclass[fleqn,10pt]{wlscirep}

\usepackage[utf8]{inputenc}
\usepackage[T1]{fontenc}

\usepackage{amsmath,amssymb,amsfonts}
\usepackage{graphicx}
\usepackage{multirow}
\usepackage[ruled,vlined]{algorithm2e}
\usepackage{textcomp}

\title{Ensemble-based graph representation of fMRI data for cognitive brain state classification}

\author[1,*]{Daniil Vlasenko}
\author[1]{Vadim Ushakov}
\author[1,2]{Alexey Zaikin}
\author[1]{Denis Zakharov}

\affil[1]{Institute for Cognitive Neuroscience, University Higher School of Economics, Moscow, 101000, Russia}
\affil[2]{Department of Mathematics and Institute for Women's Health, University College London, London, WC1H 0AY, UK}

\affil[*]{dvlasenko@hse.ru}
\begin{abstract}
fMRI is a non-invasive technique for investigating brain activity, offering high-resolution insights into neural processes. Understanding and decoding cognitive brain states from fMRI depends on how functional interactions are represented. We propose an ensemble-based graph representation in which each edge weight encodes state evidence as the difference between posterior probabilities of two states, estimated by an ensemble of edge-wise probabilistic classifiers from simple pairwise time-series features. We evaluate the method on seven task-fMRI paradigms from the Human Connectome Project, performing binary classification within each paradigm. Using compact node summaries (mean incident edge weights) and logistic regression, we obtain average accuracies of 97.07–99.74\%. We further compare ensemble graphs with conventional correlation graphs using the same graph neural network classifier; ensemble graphs consistently yield higher accuracy (88.00–99.42\% vs 61.86–97.94\% across tasks). Because edge weights have a probabilistic, state-oriented interpretation, the representation supports connection- and region-level interpretability and can be extended to multiclass decoding, regression, other neuroimaging modalities, and clinical classification.
\end{abstract}

\begin{document}

\flushbottom
\maketitle
\thispagestyle{empty}

\section*{Introduction}
One of the central objectives of modern interdisciplinary science is to study the fundamental principles underlying human brain functioning. Nowadays, the primary approach involves studying brain activity through various neuroimaging techniques, such as electroencephalography (EEG), magnetoencephalography (MEG), and functional magnetic resonance imaging (fMRI), to identify functional brain networks responsible for specific cognitive processes. Importantly, previous works have shown that the choice of neuroimaging data representation, whether at the sensor, source, or graph level, can significantly affect the outcome of brain-state classification tasks across different modalities~\cite{craik_deep_2019, saranskaia_aimbased_2025, vaghari_late_2022}. Understanding how different brain regions interact and coordinate their activity during various cognitive tasks, as well as the ability to accurately classify brain states, can provide significant insights into the nature of cognitive processes and aid in the development of diagnostic and therapeutic methods for neurodegenerative diseases. Taking into account that fMRI is a highly powerful and modern tool for investigating brain activity, offering non-invasive high-resolution insights into neural processes in vivo, it has become a key tool for mapping brain activity and analyzing cognitive states~\cite{heeger_what_2002, logothetis_what_2008}.

Analysis of fMRI data is a challenging task due to its high dimensionality and dynamic nature. In recent years, increasing attention has been devoted to applying network-based methods to represent functional connectivity between different brain regions (e.g., see~\cite{wang_graphbased_2010, richiardi_decoding_2011, takerkart_graphbased_2014}). These methods describe the brain activity using  a functional network, where nodes represent brain regions, and edges correspond to functional connections between them. This approach enables a more comprehensive understanding of cognitive states by identifying both local and global connectivity patterns in brain activity. Moreover, using network-based methods for fMRI data analysis allows not only for the investigation of brain network characteristics but also for exploring their topology with machine learning techniques including graph neural networks (GNNs)~\cite{saeidi_decoding_2022, li_graph_2019, li_braingnn_2021, bessadok_graph_2023}, which is difficult to achieve using classical statistical methods.

Currently, there are several methods for representing fMRI data in network form. The simplest and most widely used method is correlation-based graphs~\cite{li_braingnn_2021, saeidi_decoding_2022, gorban_dynamic_2021} that use Pearson's correlation coefficients as edge weights. In general, the functional connectivity between two brain regions can be computed using any metric that may or may not be invariant with respect to time and spatial positioning of brain regions~\cite{ursino_transfer_2020, hlinka_functional_2011, roebroeck_mapping_2005}. In our previous work~\cite{vlasenko_ensemble_2024}, we introduced the idea of using ensemble learning~\cite{ganaie_ensemble_2022, mohammed_comprehensive_2023, galar_review_2012} as a principled way to construct connectivity graphs from neuroimaging data. Here, we turn that concept into a fully specified and reproducible pipeline: we provide a step-by-step description of the graph-construction procedure, implement it, and empirically validate it  on task-fMRI data. 

Neuroimaging signals are noisy and heterogeneous across subjects, and individual connectivity metrics capture complementary aspects of inter-regional interactions. The key motivation of our work is to overcome these difficulties. By framing graph construction as an ensemble of simple edge-wise probabilistic models, we can (i) improve robustness to noise and inter-subject variability through averaging/aggregation, and (ii) integrate multiple time-series and connectivity-derived features into a single weighted graph that is directly optimized for the downstream binary classification objective. In this way, the resulting representation provides an informative, task-oriented inductive bias for subsequent classifiers, rather than relying on a single hand-crafted connectivity measure. Moreover, by combining information from multiple metrics into one graph, the approach avoids storing and processing several alternative graphs, which can reduce the memory usage and the computational cost of downstream network analysis. We refer to the resulting representations as ensemble (or "synolitic") graphs, from the Greek "synolo," meaning "ensemble"~\cite{nazarenko_parenclitic_2021}.

To test the effectiveness of the method, we used the Human Connectome Project (HCP) 1200 Subject Release~\cite{elam_human_2021, essen_wuminn_2013}, which offers several key advantages. It provides very high spatial and temporal resolution fMRI data, collected from hundreds of healthy participants across thousands of sessions, which ensures robust statistical power. The database includes comprehensive datasets that combine structural, functional, and behavioral information, enabling multifaceted brain analyses. Additionally, the data are standardized and openly accessible to the scientific community, promoting reproducibility and collaboration. It should be noted that this database is a good test tool for a variety of approaches to classifying brain conditions using machine learning methods. Due to the high quality of the data (high spatial and temporal resolution), these methods make it possible to achieve high classification efficiency. Typical classification efficiency by various methods was 75-95\% (e.g., see \cite{li_braingnn_2021,saeidi_decoding_2022,zhang_deep_2022}).

Using this database, we constructed ensemble graphs and performed binary classification of task-evoked experimental conditions within each paradigm using logistic regression on node-level features. For brevity, throughout the manuscript we further refer to these contrasted task-evoked experimental conditions (e.g., 0-back vs 2-back) as ``brain states'', since the proposed method is not limited to task-fMRI decoding. Across seven cognitive tasks (working memory, gambling, motor, language, social cognition, relational processing, and emotion), classification accuracy ranged from 97.07\% to 99.74\%, indicating that even a simple model operating on ensemble-graph node features can reliably distinguish the two conditions. We further compared ensemble graphs with classical correlation graphs in a GNN setting. Using the same GNN architecture, ensemble graphs achieved average accuracies from 88.00\% to 99.42\%, whereas correlation graphs performed worse, from 61.86\% to 97.94\%. Overall, ensemble graphs yielded consistently higher performance across all seven tasks, highlighting the benefit of the proposed representation. Importantly, the main contribution of this work is the representation itself rather than a novel classifier; therefore, we deliberately employ simple and widely used downstream models (logistic regression and a standard GNN) to isolate the effect of the proposed graph construction. It is also important to note that a limitation of many machine-learning approaches is the limited neurophysiological interpretability of highly accurate classifiers. More research is needed on the neurophysiological interpretability of the ML results, and it seems that our approach may help highlight task-specific functional patterns of each classified condition.

\section*{Results}
\label{sec:results}

\subsection*{Overview of data, tasks, and graph construction}
For each of the seven HCP task-fMRI experiments, we performed binary classification between two task-defined brain states. Specifically, we contrasted: Working Memory (0-back vs 2-back), Gambling (win vs loss), Motor Activity (left arm or leg vs right arm or leg), Language Processing (story vs math), Social Cognition (random movement vs mental iterations), Relational Processing (relation vs match), and Emotion Processing (neutral vs fear). Each experiment comprised 1162 labeled samples, corresponding to 581 individuals with one sample per state (i.e., 581 graphs per class). For details on data acquisition and preprocessing, see Methods; for broader context and comparison to prior HCP decoding studies, see Discussion.

We represented each sample as a complete undirected weighted graph with 379 vertices (brain regions) and 71,631 edges. For each experiment, we constructed two graph families: (i) correlation graphs as a baseline, and (ii) ensemble graphs as our proposed representation. Graphs were built under two parcellation strategies for regional time series: (i) averaging voxel time series within each region and (ii) using the first principal component (PCA) for each region.

The idea behind our proposed network representation method is as follows. In ensemble graphs, each edge weight encodes the difference between posterior probabilities of the two brain states (1 and 2) given a set of features computed from the time series of regions $i$ and $j$:
\begin{equation}
    w_{ij} = P(2 \,|\, f_1(\boldsymbol{x_i}, \boldsymbol{x_j}), \ldots, f_k(\boldsymbol{x_i}, \boldsymbol{x_j})) - P(1 \,|\, f_1(\boldsymbol{x_i}, \boldsymbol{x_j}), \ldots, f_k(\boldsymbol{x_i}, \boldsymbol{x_j})).
    \label{eq:edge_weight}
\end{equation}
Because Eq.~\eqref{eq:edge_weight} is a difference of probabilities, $w_{ij}\in[-1,1]$: negative values indicate stronger association with state~1, positive values with state~2, and larger $|w_{ij}|$ indicate higher edge-level informativeness for classification. To calculate these probabilities, a machine learning model must be trained for each edge, resulting in an ensemble of models, hence the name of the graph. When constructing ensemble graphs, features were used that allow this implementation to be considered as a processed correlation graph.  Further implementation details are provided in Methods.

\subsection*{Classification using mean incident edge weights of ensemble graphs}
In the proposed ensemble graph, each edge weight $w_{ij}$ encodes condition evidence derived from a pair of regions, i.e., it quantifies how strongly the joint information from the two regional time series supports one task condition versus the other. To obtain a compact node-level summary of an ensemble graph, we compute for each vertex $i$ the mean incident edge weight
\begin{equation}
d_i = \frac{1}{N-1}\sum_{j\neq i} w_{ij},
\qquad N=379,
\label{eq:mean_incident}
\end{equation}
yielding a 379-dimensional feature vector for each sample. This aggregation can be interpreted as a region-wise average evidence score across all pairwise interactions of that region with the rest of the brain. Importantly, it reduces the representation from $O(N^2)$ edge weights of a complete graph to $O(N)$ node features, enabling simple downstream models to operate on one vector while still leveraging information distributed over pairwise interactions.

\begin{figure}[!h]
    \centering
    \includegraphics[width=0.5\textwidth]{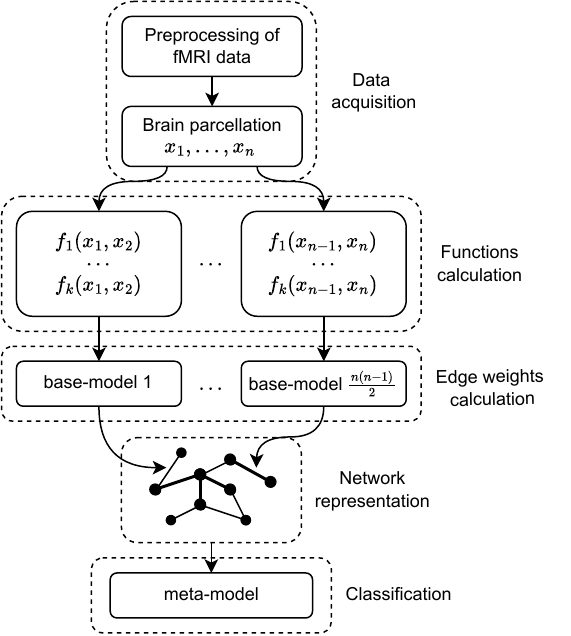}
    \caption{Pipeline of ensemble graph construction and classification.}
    \label{fig:pipline}
\end{figure}

This vector of 379 node features was then classified using a logistic regression model~\cite{pedregosa_scikitlearn_2011}. In terms of ensemble learning, graph construction models can be called base models, and graph classification model can be called meta-model. The entire pipeline, from data processing to classification, can be seen in Fig.~\ref{fig:pipline}. To evaluate the classification results and avoid information leakage between the two model levels, model validation was performed using four-fold cross-validation in accordance with the two-level training scheme described in the Methods section (Algorithm~\ref{alg:CV}). 

Across the seven experiments, this simple classifier achieved consistently high performance. With the mean parcellation strategy, average accuracy ranged from 97.07\% to 99.74\%, and with the PCA-based strategy from 96.73\% to 99.40\% (see Table~\ref{tab:res:logregr}). In all tasks, the mean-parcellation graphs yielded higher accuracy than the PCA-based graphs. 

High accuracies are consistent with the fact that we consider strongly contrasted task conditions within the HCP task-fMRI dataset. These contrasts are known to produce robust and reproducible activation differences (see Discussion); accordingly, performance above 90\% is not unexpected in this controlled within-dataset binary setting.

\subsection*{Comparison of ensemble and correlation graphs using a GNN}
Because mean parcellation provided higher accuracy in the node-summary baseline, we used only mean-parcellation method to compare ensemble and correlation representations within a GNN classifier. We trained the same GNN architecture for both graph types (see Methods) and report mean and standard deviation of accuracy across repeated runs.

Using the GNN, ensemble graphs achieved average accuracies from 88.00\% to 99.42\%, whereas correlation graphs ranged from 61.86\% to 97.94\% (Table~\ref{tab:res:gnn}). In all seven experiments, ensemble graphs yielded the best results. The mean accuracy gains (ensemble minus correlation) were largest for Emotion Processing (+26.14 pp), Gambling (+21.60 pp), Motor (+20.67 pp), and Working Memory (+17.10 pp), while Language Processing showed the smallest but stil positive gain (+1.48 pp).

\begin{figure}[!h]
  \centering
  \includegraphics[width=0.75\textwidth]{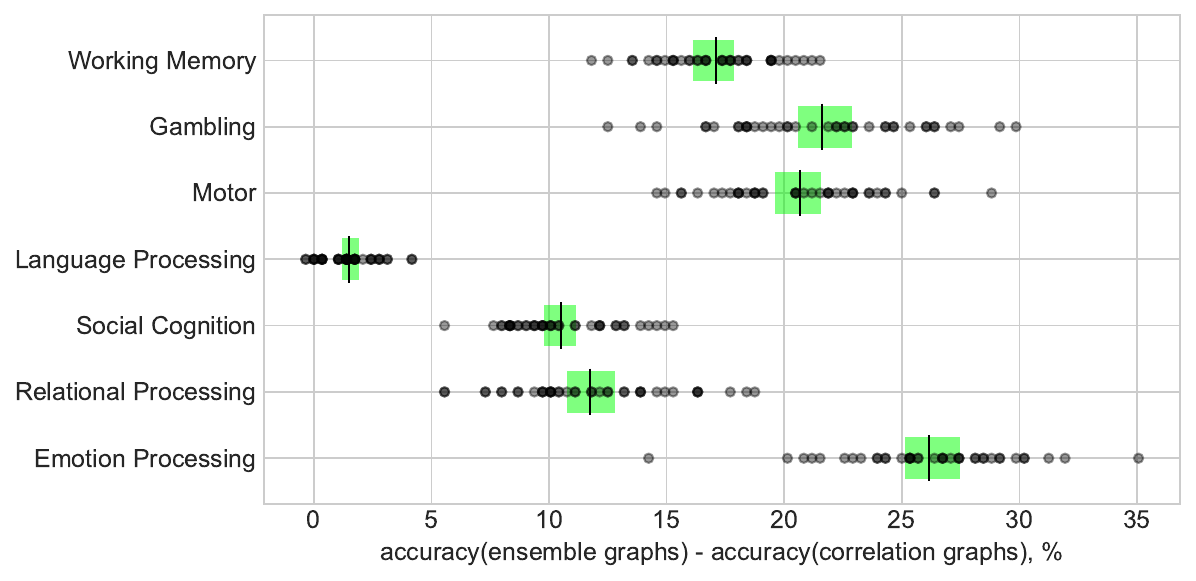}\\[1em]
  \includegraphics[width=0.75\textwidth]{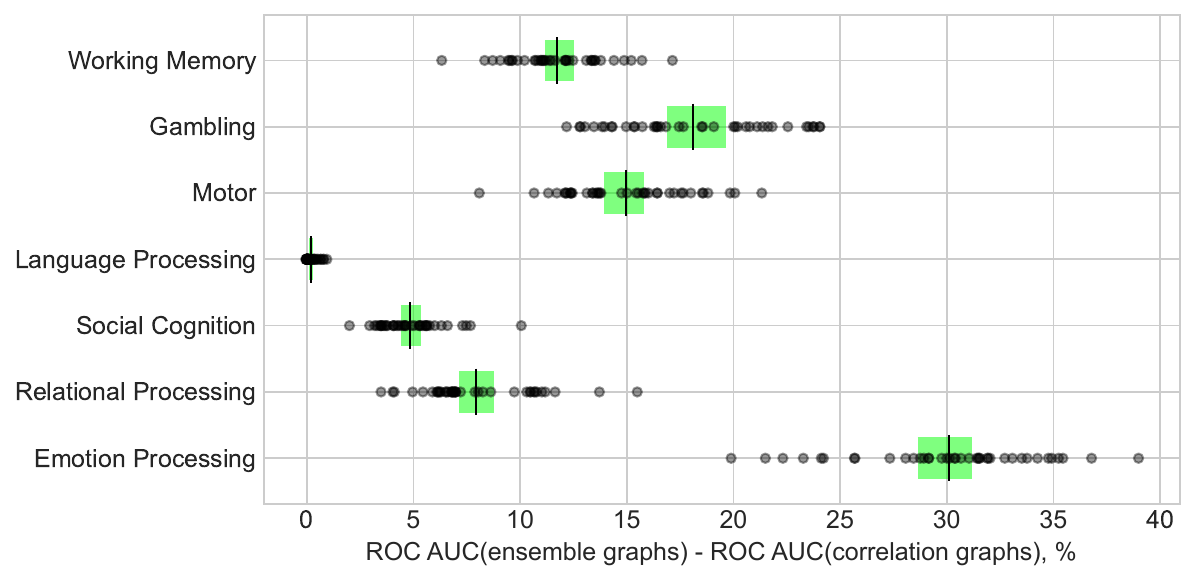}
  \caption{Bootstrap percentile-t confidence intervals for run-level metric differences between ensemble-based and correlation-based graphs.}
  \label{fig:confLevels}
\end{figure}

To quantify uncertainty in the performance gap between graph representations, we computed bootstrap percentile-$t$ confidence intervals (see Methods) over run-level metric differences (ensemble minus correlation). In Fig.~\ref{fig:confLevels}, intervals lying entirely above zero indicate statistically significant improvements for ensemble graphs.

\subsection*{Visualization of ensemble graphs}
\begin{figure}[h]
    \centering
    \includegraphics[width=0.49\textwidth]{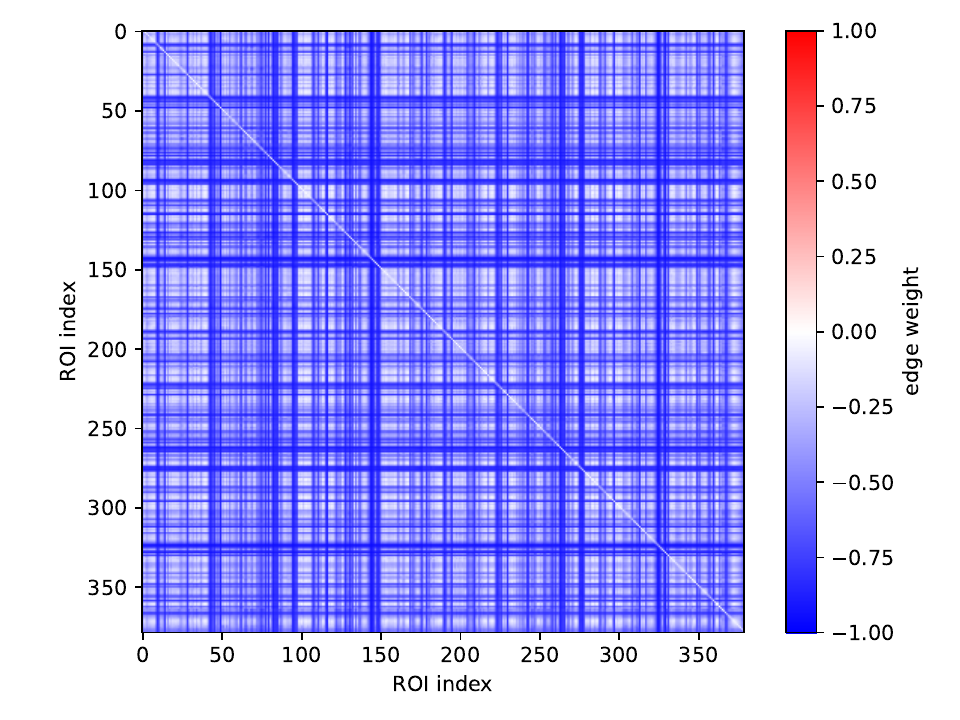} 
    \includegraphics[width=0.49\textwidth]{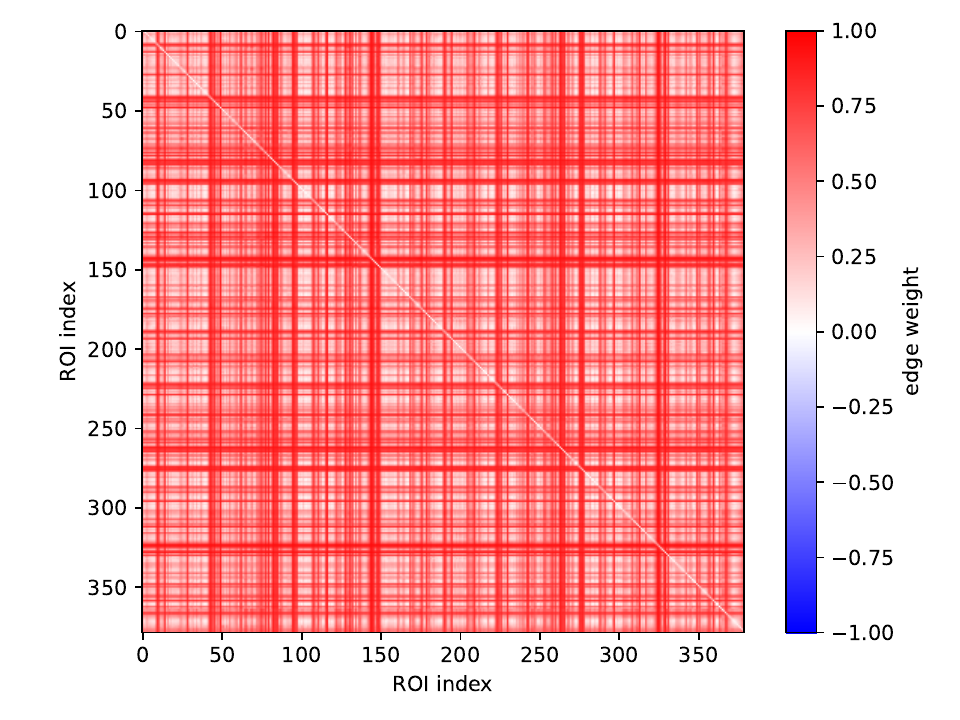} 
    \caption{Averaged edge weight matrices of ensemble graphs for ``story'' (left) and ``math'' (right) brain states in the language processing experiment.}
    \label{fig:edge_wieghts}
\end{figure}

\begin{figure}[h]
    \centering
    \includegraphics[width=0.35\textwidth]{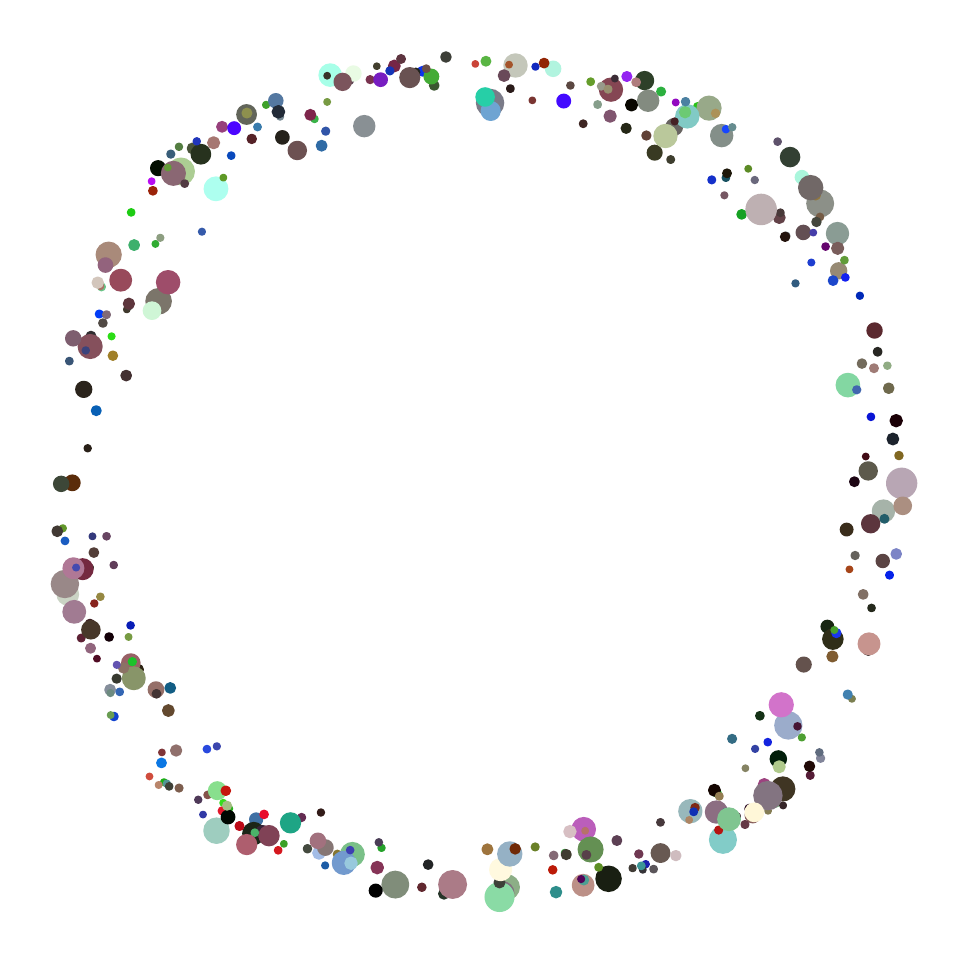} 
    \includegraphics[width=0.35\textwidth]{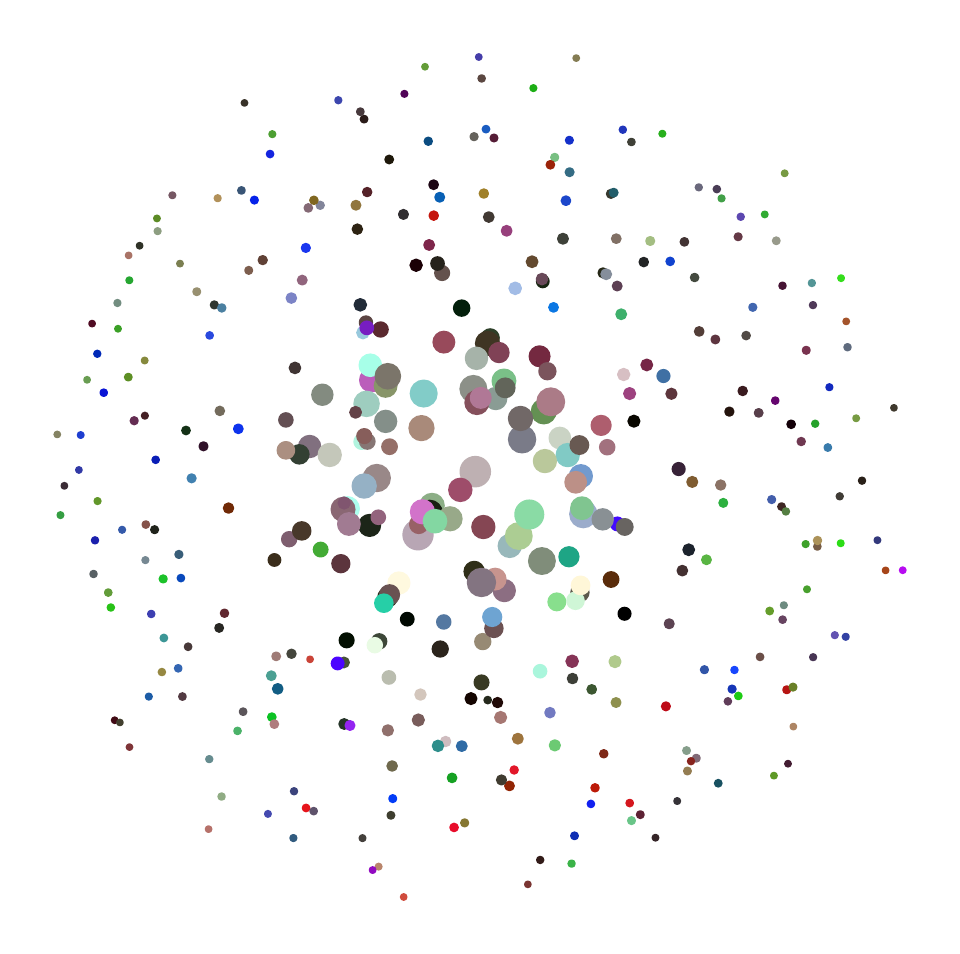} 
    \caption{Application of the Fruchterman-Reingold force-directed algorithm to the averaged edge weight matrices of ensemble graphs for ``story'' (left) and ``math'' (right) brain states in the language processing experiment.}
    \label{fig:spring_layout}
\end{figure}

Finally, we illustrate qualitative differences between brain states using visualizations of ensemble graphs (mean-parcellation only). For visualization, edge weights were averaged across individuals in the meta-level test sets obtained during cross-validation (see Methods and Algorithm~\ref{alg:CV}). We used two complementary views: (i) heat maps of averaged edge-weight matrices (Fig.~\ref{fig:edge_wieghts}) and (ii) a Fruchterman–Reingold force-directed layout~\cite{fruchterman_graph_1991} applied to the same averaged matrices (Fig.~\ref{fig:spring_layout}).

As an example, Fig.~\ref{fig:edge_wieghts} shows averaged ensemble edge-weight matrices for the two brain states (``story'' and ``math'') in the language-processing experiment. Negative edge weights indicate stronger association with state ``story'', whereas positive weights indicate stronger association with state ``math''. The force-directed layouts in Fig.~\ref{fig:spring_layout} reflect the same edge-weight information; for clarity, edges are not explicitly drawn due to the complete-graph density, but they influence node positions. Node sizes are determined as difference in mean incident edge weights of two states and then scaled (with an exponential transformation) to emphasize regional contrasts. 

In both cases of visualization it can be seen that some regions stand out as more important for classification. In the case of matrices, such regions give rise to strongly shaded lines. In the case of graphs themselves, the vertices representing such regions are large and close to the center in one of the states (when edge weights tend to be positive).

\section*{Discussion}
In this study we proposed an ensemble-based graph representation of task-fMRI data in which edge weights encode brain states via differences in posterior probabilities. The results demonstrate that ensemble graphs consistently provide a more informative network substrate than correlation graphs for binary brain-state classification across the considered HCP task experiments. Notably, high performance is obtained even when the meta-model operates on coarse node-level summaries (mean incident edge weights), indicating that state information is expressed at the level of distributed network interactions and remains recoverable after substantial dimensionality reduction (see Results; Methods for definitions).

Although the pipeline with logistic regressor achieved slightly higher accuracy than the GNN implementation we used, we attribute this primarily to sample-size limitations and the specific architecture choices. Future work will explore data-augmentation strategies to expand the dataset and perform a systematic search for an optimal GNN design tailored to ensemble graphs. Our principal aim in this study was to introduce a novel graph-based fMRI representation; detailed metamodel optimization will be the focus of subsequent investigations.

A practical advantage of the proposed formulation is interpretability. Because each edge weight has a probabilistic, state-oriented meaning, one can interpret node-level aggregates as region-wise evidence derived from pairwise contributions. This creates a natural interface to downstream analyses in network neuroscience: identifying consistently informative regions/edges, examining task-specific subnetworks, and applying community-level methods. The visualizations in Results illustrate this potential, but establishing neurobiological interpretations will require careful statistical control and validation beyond visual inspection.

Future work can proceed in several directions. Methodologically, our framework is not tied to any specific base models that calculate weighted graph: in principle, any probabilistic classifier can serve as a base model, provided it remains simple enough to train reliably on typical neuroscience sample sizes. Likewise, the proposed ensemble-graph construction is not tied to a particular functional-connectivity estimator and can be interpreted as an additional discriminative weighting step applied to graphs built from the same underlying time-series data (e.g., static, dynamic/windowed, or alternative dependence measures). Although we have focused on binary classification, this approach can be extended to multiclass settings by using multiclass base models so that there are as many edges between two vertices as there are classes in the data, then the weight of the edge will be equal to the probability of the corresponding class. Alternatively, the problem can be reformulated for regression by replacing classifiers with regression models. Because the output is a graph-structured representation, it naturally invites the use of network-analysis tools and graph-based machine learning. An important open question is the choice of an effective meta-model; promising directions include graph learning methods and explainable AI approaches~\cite{ying_gnnexplainer_2019, luo_parameterized_2020, yuan_xgnn_2020} to identify informative connections, regions, and communities. In addition to fMRI, the same idea can be adapted to EEG/MEG, for which it will be possible to construct a graph not in the space of signal sources (brain regions), but in the space of sensors. Finally, the methodology can be applied to pathological brain states, enabling classification of disease-related network alterations and potentially supporting diagnosis and monitoring of neurological disorders.

\subsection*{Relation of our HCP data classification task to the existing literature}
Network-based classification on HCP task-fMRI has been explored previously, typically using correlation-derived graphs and GNN architectures. For example, Li et al.~\cite{li_braingnn_2021} proposed an interpretable BrainGNN with region selection and reported 94.4\% average accuracy across seven HCP experiments. Similarly, Saeidi et al.~\cite{saeidi_decoding_2022} applied correlation-based representations with a three-layer GNN augmented by node-embedding techniques (e.g., NetMF, Node2Vec), achieving 97.7\% with optimal hyperparameters. In a broader setting, Zhang et al.~\cite{zhang_deep_2022} addressed multistate classification (21 states) by constructing a resting-state network and overlaying task-fMRI dynamics, reaching 89.8\% mean accuracy. 

In contrast, our work uses only task-fMRI and targets binary classification separately within each of the seven HCP experiments. Our primary goal is not to optimize performance under alternative problem formulations, but to introduce a novel ensemble-based graph construction and demonstrate its effectiveness with minimal complexity. We therefore avoid extensive meta-model engineering and report results with simple, widely used models to emphasize that the performance gains arise from the representation. We intentionally chose HCP task-fMRI as a large, standardized benchmark with high data quality and multiple paradigms, enabling a controlled methodological comparison. Although we use only HCP data, we evaluate seven distinct paradigms/tasks, which provides a broad range of decoding settings and reduces the risk that the results are specific to a single contrast.

\section*{Methods}
\label{sec:methods}

\subsection*{Data and preprocessing}
We used task-fMRI data from the Human Connectome Project (HCP) 1200 Subject Release.
From this release, we selected 581 healthy participants (295 female, 286 male), aged 22--35 years,
with no reported anatomical abnormalities and complete data for all seven HCP task-fMRI experiments
considered in this study~\cite{elam_human_2021, essen_wuminn_2013, barch_function_2013}.

Task-fMRI was acquired on a 3T MRI scanner (field of view $208\,\mathrm{mm}\times 180\,\mathrm{mm}$;
$2.0\,\mathrm{mm}$ isotropic voxels; TR $=0.72\,\mathrm{s}$; TE $=33.1\,\mathrm{ms}$; flip angle $=52^{\circ}$).
To maintain consistency across subjects, we used data with the right-to-left phase encoding direction only.
Preprocessing followed the HCP minimal preprocessing pipeline (including motion correction, distortion correction, spatial normalization, and noise filtering); details are provided in the HCP documentation and primary pipeline publications~\cite{barch_function_2013, glasser_minimal_2013}. In addition to the HCP pipeline, we applied linear trend removal and z-scoring of voxel time series prior to
parcellation.

We parcellated the brain into 379 regions using the HCP multi-modal parcellation (MMP) atlas
(180 cortical regions per hemisphere plus 19 subcortical regions)~\cite{glasser_multimodal_2016}. For each region, we constructed two alternative representative regional time series (analyzed as two separate
datasets): (i) voxelwise averaging within each parcel, and (ii) the first principal
component of voxelwise signals within the parcel.

For each of the seven HCP task-fMRI experiments, we defined two task-contrast brain states and performed binary classification within each experiment (one sample per state per individual, yielding 1162 labeled samples per experiment). Thus, we obtained seven binary classification tasks: Working Memory (0-back vs 2 back), Gambling (win vs loss), Motor Activity (left arm or leg vs right arm or leg), Language Processing (story vs math), Social Cognition (random movement vs mental iterations), Relational Processing (relation vs match), and Emotion Processing (neutral vs fear).

\subsection*{Graph representation}
We defined a complete undirected weighted graph $g = (V, E, H, W)$, where $ V = \{i \,|\, i \in 1, \ldots, n\} $ represents the set of vertices, $ E = \{ij \,|\, i \in V; j \in V; i \neq j\} $ represents the set of undirected edges, $ H = \{\boldsymbol{h_i} \,|\, i \in V\} $ denotes the vectors of values associated with vertices, and $ W = \{w_{ij} \,|\, ij \in E\} $ denotes the weights of the edges. Each vertex corresponds to a brain region derived from the parcellation of fMRI data. The time series of a brain region~$i$ is denoted by~$\boldsymbol{x_i}$. 

\subsubsection*{Correlation graphs}
For correlation graphs, the node feature vector for region $i$ was defined as $h_i=(\mathrm{mean}(x_i), \mathrm{std}(x_i))$, where $x_i$ is the regional time series. Edge weights were Pearson correlations $w_{ij}=\rho(x_i,x_j)$. For comparability across samples, edge weights were z-scored within each graph. We emphasize that our goal is not to benchmark alternative functional-connectivity estimators (static vs dynamic FC, or different dependence measures). Instead, we keep the input data and the simplest widely used static correlation-based graph as a fixed reference representation, and ask whether the proposed ensemble-based weighting provides a more informative graph substrate for downstream decoding under identical conditions.

\subsubsection*{Ensemble graphs}
For ensemble graphs, edge weights were computed as the difference between posterior probabilities of the two brain states, as defined in Eq.~\eqref{eq:edge_weight} (introduced in Results). To implement Eq.~\eqref{eq:edge_weight}, we used a compact feature set $ f_1, \ldots, f_k $ composed of simple statistics of the two regional time series: the mean and standard deviation of each region's time series and the Pearson correlation between the two time series. Pearson correlations were z-scored (as for correlation graphs) before being used as an input feature. Computing each $w_{ij}$ in Eq.~\eqref{eq:edge_weight} requires a probabilistic classifier trained to predict the brain state from the edge-level feature vector. We trained one classifier per edge. In this study, each base model was logistic regression (scikit-learn) with default settings (regularization $C=1.0$, $\ell_2$ penalty, lbfgs solver, maximum 100 iterations)~\cite{pedregosa_scikitlearn_2011}. For the GNN experiments, node features were set to a constant $h_i=1$, since the edge weights of ensemble graphs already contain encoded information from the brain rigeons.

We emphasize that our goal is not to benchmark different functional-connectivity estimators. We use standard static correlation graphs as a fixed reference and evaluate whether the proposed ensemble-based weighting yields a more informative graph representation under the same input data and preprocessing.

\subsection*{Cross-validation and leakage control}
The cross-validation process was designed to account for the two levels of training in our pipeline: the base-models (logistic regressions) used to construct the graphs and the meta-models (logistic regression or a GNN) used for classification. Below is a detailed description of the algorithm we devised~(Algorithm~\ref{alg:CV}).

To begin, the parceled fMRI dataset was split into four folds with subject-level grouping. Specifically, for each participant, the two samples corresponding to the two contrasted states were treated as a paired unit and were always assigned to the same fold, preventing leakage of subject-specific signatures across training and test sets. We then used a nested cross-validation scheme. In the outer loop, each fold $G_i$ was selected in turn as the held-out test set for the meta-model, while the remaining three folds formed the outer training set. Within this outer training set, an inner loop was used to train the base models that produce edge weights and to construct graphs in an out-of-fold manner. Concretely, for each inner fold $G_j$ among the three outer training folds, the base models were trained on the parceled fMRI data from the other two folds and then used to construct graphs for all samples in $G_j$. After completing the inner loop, the base models were retrained on the parceled fMRI data from all three outer training folds (for which graphs had already been constructed) and were then used to construct graphs for the outer test fold $G_i$. Finally, the meta-model was trained on the graphs from the three outer training folds and evaluated on the graphs from the held-out fold $G_i$.

This cross-validation scheme allows for the most efficient use of the dataset across two consecutive levels of model training without data leakage between the training and test samples.

\subsection*{Meta-models for classification}

\subsubsection*{Logistic regression on node summaries}
To obtain a compact graph-level representation, we computed the mean incident edge weight for each vertex
(Eq.~\eqref{eq:mean_incident}, introduced in Results), producing 379 node-level features per sample. These features were classified using logistic regression with default scikit-learn parameters~\cite{pedregosa_scikitlearn_2011}. 

\subsubsection*{Graph neural network for graph-level classification}
\begin{figure}[!h]
    \centering
    \includegraphics[width=0.45\textwidth]{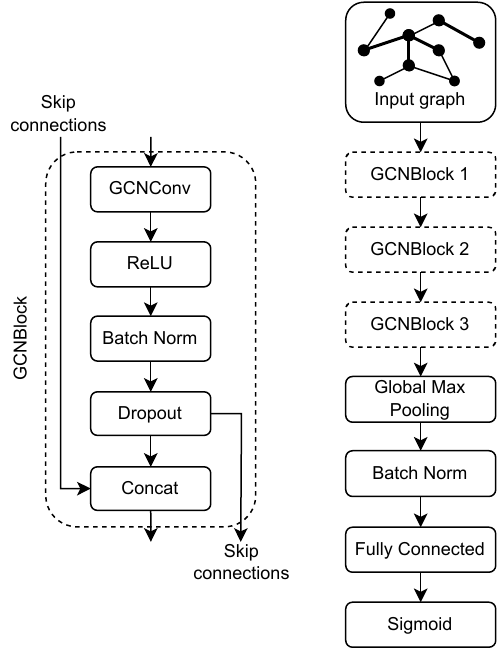}
    \caption{Architecture of the GNN used to compare the efficiency of ensemble and correlation graphs in a classification task.}
    \label{fig:GNN}
\end{figure}

We employ a GNN built from repeating units we call GCNBlocks (Fig.~\ref{fig:GNN}). Each GCNBlock begins with a graph convolutional layer (GCNConv), which aggregates information from node’s neighbors according to the graph adjacency structure~\cite{kipf_semisupervised_2017}. Immediately following convolution, we apply a ReLU activation~\cite{nair_rectified_2010} and Batch Normalization~\cite{ioffe_batch_2015} to improve convergence and stabilize training. A Dropout layer then randomly deactivates a fraction of the feature channels, reducing overfitting by preventing the network from relying too heavily on any single neuron~\cite{srivastava_dropout_2014}. To preserve and fuse information across depths, each GCNBlock incorporates skip connections~\cite{he_deep_2015}. Within each GCNBlock we concatenate the block’s new features with the previous block’s Dropout output (or, for the first block, with the original node features): specifically, GCNBlock 1 fuses its post-dropout features with the original input node features; GCNBlock 2 fuses its post-dropout features with GCNBlock 1’s post-dropout features; and GCNBlock 3 fuses its post-dropout features with GCNBlock 2’s post-dropout features.

We stack three GCNBlocks in sequence. After the final block, a global max-pooling operation~\cite{xu_how_2019} reduces all node embeddings into a single, fixed-length vector by taking the maximum activation over nodes for each feature channel. This pooled graph representation is then passed through a Batch Normalization layer, ensuring that its distribution remains well-conditioned before entering the classification head. A fully connected layer maps the normalized embedding into the target prediction space, and a sigmoid activation produces a probability in [0,1] for binary classification. Training is driven by the binary cross-entropy loss.

To optimize performance without over-smoothing node representations, we found that three convolutional blocks strike the right balance: they extend the receptive field beyond immediate neighbors while avoiding excessive feature homogenization. During training, we used an Adam optimizer~\cite{kingma_adam_2017} with hyperparameters determined on a small hold-out set. To further enhance convergence and combat overfitting, we apply a LinearLR scheduler that linearly decays the learning rate over the total number of iterations. The network was trained for each experiment out of seven possible ones and for each graph type out of two possible ones separately. During GNN training and evaluation, we used paired sampling within each fold: whenever a subject's graph from one state was included in a mini-batch, the corresponding graph of the same subject from the other state was also included. This pairing was applied strictly within the predefined subject-level folds described above.

For model validation, we performed four-fold cross-validation~(see Algorithm~\ref{alg:CV}). In each fold, the network was trained for 100 epochs, and to account for randomness we repeated this process 10 times with different random seeds. From these runs we compute mean values and standard deviations of Accuracy and ROC-AUC. To quantify uncertainty in the difference between our ensemble-based graphs and traditional correlation-based graphs, we build percentile-t confidence intervals by drawing 100 bootstrap samples from the 40 run-level metric differences~\cite{efron_introduction_1994}.

\subsection*{Graph visualization}
We visualized ensemble graphs using (i) heat maps of averaged edge-weight matrices and (ii) a 2D force-directed layout (Fruchterman--Reingold) implemented in NetworkX with default parameters~\cite{fruchterman_graph_1991}. Visualizations were computed using edge weights averaged across individuals in the meta-test sets of cross-validation; for clarity, edges were not drawn in the force-directed plots due to the complete-graph density. Colors of the vertices correspond to the colors of brain regions as defined by the MMP atlas~\cite{glasser_multimodal_2016}. The size of each vertex is determined as follows. The difference between the mean incident edge weights of a vertex in the two brain states is first calculated. This difference is then scaled by a multiplier $z$ and raised to an exponential power. The scaling ensures that the largest vertex size matches the maximum value of 200 for the node\_size parameter in draw\_networkx\_nodes function. The exponential transformation highlights differences more distinctly, making variations between regions visually clearer. 

\subsection*{Implementation and reproducibility}
All graph construction, cross-validation, and model training/testing code (including logistic regression and GNN experiments) is available in the accompanying repository (see Data availability).

\bibliography{SciRep}

\section*{Author contributions statement}
D.V. designed the study, developed the ensemble-based graph construction method, performed all non-GNN experiments, analysed the results, and wrote the manuscript. V.U. implemented, trained, and evaluated the GNN models and contributed to the interpretation of the GNN results. A.Z. and D.Z. contributed to the study design and interpretation of the results. D.Z. also contributed to the supervision and conceptualization. All authors reviewed and approved the manuscript.

\section*{Acknowledgements}
The research was supported by the Russian Science Foundation (grant number 24-68-00030) and in part through computational resources of HPC facilities at NRU HSE~\cite{kostenetskiy_hpc_2021}

\section*{Data availability}
The data analysed in this study are from the Human Connectome Project (HCP) 1200 Subject Release and are available via ConnectomeDB and BALSA (subject to the HCP data use terms).  The code that constructs ensemble graphs, implements cross validation and performs training and testing of meta-models, including logistic regression and graph neural network, is available at \href{https://github.com/Daniil-Vlasenko/EnsGraph2025.git}{https://github.com/Daniil-Vlasenko/EnsGraph2025.git}.

\section*{Additional information}

\noindent\textbf{Accession codes.} Not applicable.

\noindent\textbf{Competing interests.} The authors declare no competing interests.

\begin{algorithm}[h]
\SetAlgoLined
\KwIn{Dataset $D$, Number of groups $G = 4$}
\KwOut{Performance metrics}
\vspace{0.3cm}

Split $D$ into $G$ subject-level folds: $D_1, D_2, \ldots, D_G$ (paired samples from each subject are kept within one fold);

\ForEach{$i \in \{1, 2, \ldots, G\}$}{
    $D_i \leftarrow \text{test set for meta-model}$\;
    $D_{\text{meta-train}} \leftarrow \{D_j \mid j \neq i\}$\;
    
    \ForEach{$j \in \{1, 2, \ldots, G\} \setminus \{i\}$}{
        $D_j \leftarrow \text{set for graph construction}$\;
        $D_{\text{base-train}} \leftarrow \{D_k \mid k \neq i, k \neq j\}$\;
        
        Train base-models on $D_{\text{base-train}}$\;
        Compute graphs for $D_j$ using the trained base-models\;
    }
    
    Retrain base-models on $D_{\text{meta-train}}$\;
    Compute graphs for $D_i$ using the retrained base-models\;
    Train meta-model on graphs from $D_{\text{meta-train}}$\;
    Test meta-model on graphs from $D_i$\;
}
\caption{Cross-Validation Procedure}
\label{alg:CV}
\end{algorithm}

\begin{table}[!h]
\centering
\caption{Classification results in the format ``mean (standard deviation)'' based on mean incident weights of vertices with logistic regression for seven tasks and two parcellation methods.}
\begin{tabular}{lccc}
\hline
 & & Accuracy, \% & ROC AUC, \% \\ \hline
\multirow{2}{*}{Working Memory} & mean & \textbf{98.19 (0.86)} & \textbf{99.88 (0.03)} \\
 & pca  & 97.42 (0.58)         & 99.78 (0.04)          \\ \hline
\multirow{2}{*}{Gambling}       & mean & \textbf{97.07 (0.39)} & \textbf{99.20 (0.32)} \\
 & pca  & 96.73 (0.72)         & 99.19 (0.46)          \\ \hline
\multirow{2}{*}{Motor Activity} & mean & \textbf{99.57 (0.28)} & \textbf{99.99 (0.01)} \\
 & pca  & 98.54 (0.95)         & 99.96 (0.04)          \\ \hline
\multirow{2}{*}{Language Processing} & mean & \textbf{99.74 (0.29)} & \textbf{100.00 (0.01)} \\
 & pca  & 99.31 (0.42)         & 99.99 (0.01)           \\ \hline
\multirow{2}{*}{Social Cognition} & mean & \textbf{99.48 (0.17)} & \textbf{100.00 (0.00)} \\
 & pca  & 99.40 (0.29)         & 99.99 (0.01)           \\ \hline
\multirow{2}{*}{Relational Processing} & mean & \textbf{98.11 (0.52)} & \textbf{99.89 (0.08)} \\
 & pca  & 97.68 (1.20)         & 99.81 (0.12)           \\ \hline
\multirow{2}{*}{Emotion Processing} & mean & \textbf{98.45 (1.04)} & \textbf{99.76 (0.23)} \\
 & pca  & 97.42 (0.99)         & 99.65 (0.25)           \\ \hline
\end{tabular}
\label{tab:res:logregr}
\end{table}

\begin{table}[!h]
\centering
\caption{Classification results in the format ``mean (standard deviation)'' based on GNN for correlation and ensemble graphs for mean parcellation method.}
\begin{tabular}{llcc}
\hline
 &  & Correlation & Ensemble \\ \hline
\multirow{2}{*}{Working Memory} & Accuracy, \% & 80.36 (2.06) & \textbf{97.46 (1.46)} \\
 & ROC AUC, \% & 88.02 (2.16) & \textbf{99.73 (0.30)} \\ \hline
\multirow{2}{*}{Gambling} & Accuracy, \% & 72.27 (3.90) & \textbf{93.87 (1.56)} \\
 & ROC AUC, \% & 79.98 (3.51) & \textbf{98.11 (0.77)} \\ \hline
\multirow{2}{*}{Motor} & Accuracy, \% & 77.96 (3.29) & \textbf{98.63 (1.11)} \\
 & ROC AUC, \% & 84.97 (2.83) & \textbf{99.92 (0.12)} \\ \hline
\multirow{2}{*}{Language Processing} & Accuracy, \% & 97.94 (1.23) & \textbf{99.42 (0.51)} \\
 & ROC AUC, \% & 99.78 (0.26) & \textbf{99.98 (0.06)} \\ \hline
\multirow{2}{*}{Social Cognition} & Accuracy, \% & 88.40 (2.07) & \textbf{98.90 (0.63)} \\
 & ROC AUC, \% & 95.07 (1.50) & \textbf{99.93 (0.11)} \\ \hline
\multirow{2}{*}{Relational Processing} & Accuracy, \% & 83.17 (2.69) & \textbf{94.90 (2.24)} \\
 & ROC AUC, \% & 91.33 (2.52) & \textbf{99.25 (0.50)} \\ \hline
\multirow{2}{*}{Emotion Processing} & Accuracy, \% & 61.86 (3.45) & \textbf{88.00 (2.40)} \\
 & ROC AUC, \% & 65.01 (4.24) & \textbf{95.12 (1.77)} \\ \hline
\end{tabular}
\label{tab:res:gnn}
\end{table}

\end{document}